\documentclass[proceedings]{JHEP3}
\PrHEP{PrHEP JHW2002}
\conference{Twenty-sixth Johns Hopkins Workshop}

\newcommand{\BEQ}{\begin{equation}}
\newcommand{\EEQ}{\end{equation}}

\usepackage{epsfig}

\title{What Can We Learn from Cosmic Rays?}
\author{\speaker{S.~K\"{o}vesi-Domokos}
and G.~Domokos\\
Department of Physics and Astronomy\\
The Johns Hopkins University\\
E-mail: \email{skd@jhu.edu}}
\abstract{Ultra high energy cosmic rays (UHECR) pose a problem
either for particle physics or for astrophysics (or for both)
by the unexpectedly high number of cosmic ray showers observed 
with energy above  $ \approx 5 \times 10^{19}$ eV , the Greisen-Zatsepin-Kuzmin
(GZK) cutoff~\cite{gzk}. Our emphasis is on those possible solutions
of the puzzle which assume that ultra high energy neutrinos travel
cosmic distances. We present, in detail, a model  which is based on 
a low energy (50 to 100 TeV) transition to a higher than four
dimensional string regime.  Neutrino-quark cross sections grow
exponentially close to the threshold of this new scale because of 
the fast increase of the density of string states and effectively acquire 
hadronic strength.}          
\begin{document}
\section{Introduction}
 Cosmic rays were discovered about ninety years ago by Hess and
Kohlh\"{o}rster \cite{hess}. Through all these years
they have provided a rich source of vital information on our galaxy
and the universe beyond to astrophysics. Until the early nineteen fifties
experimental particle physics could use only cosmic rays as the ``incoming
beam'' for creating members of the particle zoo never seen before, like 
muons, pions, kaons, the lambda {\em etc.} However, the emerging accelerator
physics with its well-controlled experimental environment put the interest
in cosmic ray observations to the back burner as far as particle physics 
was concerned. During the last decade or so a dramatic turn-about took place.
Atmospheric and solar neutrino detectors gave the first hint to the
somewhat clueless particle physics community on the physics beyond the
hugely successful, nevertheless incomplete  Standard Model. Reactor 
and accelerator neutrino experiments gave confirmation of those cosmic
ray results just last year.   The limit on cosmic diffuse
gamma radiation background by \mbox{EGRET}~\cite{egret} combined with measurements of the
cosmic ray spectrum at the highest energies  provides restrictions
on possible  grand unified theories. The number of deeply penetrating horizontal
showers measured by the grand array, \mbox{AGASA}~\cite{agasa} and the nitrogen fluorescence
detector Fly's Eye \cite{hires} supplies  bounds 
on larger than four dimensional models with low scale gravity \cite{anchor}.
These are just some examples for the renewed interest in using
observations concerning  astrophysical processes and properties of the hot big bang
cosmology as possible tests for predictions of
theories going beyond the Standard Model~\cite{raffelt}.

 \section{The significance of the high energy end of the spectrum}

  A simplistic characterization of the  measured
cosmic  ray spectrum affirms its isotropy, a composition, which is overwhelmingly hadronic
(protons and nuclei) and a power law ($ \propto E^{-\alpha} $)
 energy dependence  through about ten
orders of magnitude from $10^{10}$~eV  to $10^{20}$~eV. There are two
energy regions, $\approx 4 \times 10^{15}$ eV and $\approx 10^{18.5}$ eV (named
 the ``knee'' and the ``ankle'', respectively)  where $\alpha$ changes quite abruptly:
$\alpha \approx 2.7$ for $E  \lesssim  10^{15.5}$ eV,  $\alpha \approx 3.1$
 for $10^{15.5}$eV$ \lesssim E\lesssim 10^{18.5}$eV  and
 $\alpha \approx 2.8.$ for $E\gtrsim 10^{18.5}$eV. For more details
see~\cite{rebel}. Although a considerable number of
substantially improved old  detectors as well as new ones will provide  
crucial experimental observations for the ``knee area'' in cosmic
rays~\cite{rebel}, for GeV and TeV gamma-rays  and -- hopefully -- for neutrinos 
in the near future,  the present paper brings into focus only the ultra high
energy region of the cosmic ray spectrum.

 The CM energy per particle at the Tevatron and at
the LHC  (1 TeV and 7  TeV, respectively) fall  between 
the knee and the ankle on the cosmic ray spectrum,
 corresponding to fixed target energies
  $\approx 10^{15}$ eV and $\approx 10^{17} $~eV.  The highest energy
event ever recorded has energy  $3\times 10^{20}$~eV (by the Fly's Eye detector~\cite{fly}),
which is equivalent to about 500~TeV in the CM system;  otherwise, its extensive 
air shower (EAS) fits the properties of 
a proton or maybe a nucleus induced shower~\cite{halzen}. Clearly, at the high energy end of the
spectrum the primary and  some of the secondary interactions at the top of the atmosphere
involve processes never tested in accelerator experiments;  moreover, the highest
energy events certainly scrutinize  physics beyond the electroweak scale.
 
 In addition, the region above the ankle presented the GZK puzzle. 
Up to about $10^{19}$~eV energy the universe is transparent for protons, although
the interaction with the everywhere present regular and chaotic magnetic fields
takes away the directional information pointing back to their source.
However, above that energy  protons
produce $e^{+}e^{-}$ pairs and more importantly pions\footnote{Around this energy
and above it the effect of the deflection by the galactic and intergalactic magnetic fields
is small enough that with reasonable certainty a ``source box''of a few degrees
can be defined even for charged
incoming cosmic particles.}on the cosmic microwave background (CMB) radiation .
 At the energy $10^{19.5}$ eV
the mean free path for protons is down to about 7-8~Mpc and at each collision
they lose about 20 -- 25~\% of their energy. Consequently, the expectation~\cite{gzk}  was
that there should be a cutoff and just under the cutoff a pileup (a bump on the spectrum)
as a consequence of the pion photoproduction for the above threshold protons or 
nuclei~\cite{hill}. (Nuclei photodisintegrate on the CMB and photons are absorbed
(through $e^+ e^-$ production) on the CMB and on the universal radio background.)

There are somewhat less than a hundred extensive
air showers observed with energies above $ 10^{19.5} $~eV and about ten events above
$ 10^{20}$~eV. The arrival directions are compatible with an isotropic distribution 
of sources. Wherever determination was possible, the properties of the primary particles
starting the EAS were consistent with being   hadrons.

 There is considerable disagreement on the errors of EAS energy estimates, on the finer
features of the spectrum, and on the statistical significance of the data
among the  collaborations \cite{volcano, havera,
yakutsk, fly, agasa}, which provided the observations. The recent analysis of the 
data already on file, the improvement of statistics by the continued operation
of AGASA and HiRes, and  by the future observations of new
detectors (under construction like the
Pierre Auger Observatory \cite{auger}, and the planned Owl/Airwatch \cite{owl}) 
the reliability of
the observational data will substantially improve in the next couple of years.

 The propagation
of nucleons, nuclei and gamma rays was studied in detail \cite{sigl} with the conclusion that
their sources cannot be very distant (at most $\approx 50 - 70$ Mpc away)\footnote{If the effect
of the random extragalactic magnetic fields are taken into account, in addition
to the energy loss processes,  the average GZK distance of a  proton further decreases
\cite{stanev} \cite{domokos}.}. There are only a few astrophysical sources, which can 
possibly accelerate particles to these extreme energies ({\em e.g.} AGN-s, radio galaxies). 
Since few of these objects are in the "GZK sphere" and the requirement of isotropy of the incoming
extreme energy cosmic rays (EECR) also presents a severe restriction on their origin,
the community started looking for new explanations for the higher than expected number of  
particles  at the very end of the spectrum.

\section{GZK evading messengers}

There are two large classes of models which can provide interpretation for the
data, {\em top down} and  {\em bottom up} models\footnote{For completeness, we must add the possibility
of Lorentz symmetry violation, which could eliminate the GZK cutoff completely \cite{
batsigl, bhatta}.}.

{\em Top down} models~\cite{batsigl, bhatta} all use physics beyond the Standard Model.
Topological defects and long lived 
relic particles could be produced  everywhere during  phase transitions experienced by the
Universe after the Big Bang. Through their collapse/decay they produce the small number of EECR
(protons) and dominantly photons and neutrinos.Their major advantage
is that they avoid the problems of long range propagation in the background photon "gas", as well as
the difficulties of accelerating particles to extreme energies. The art in the construction of these
models is to make the required 
flux of EECR without creating too large a contribution to the diffuse gamma ray background 
bounded by the results of the \mbox{EGRET detector}~\cite{egret}. Horizontal EAS (zenith angle larger than
$ 60^\circ $) give limits for the neutrino flux.  
 
{\em Bottom up} scenarios all face the acceleration problem. However, if --  as a first step -- 
one can assume that the sources can be outside of the  GZK sphere
 ($R_{GZK} \approx$  couple of tens Mpc), then the number of sites able to produce EECR
increases considerably. Among  Standard Model particles only the neutrino can handle 
cosmic distances without dramatic absorption\footnote{In principle,
some supersymmetric particles can have their GZK cutoff at higher energies than the nucleons.
However, LEP2 already excluded light superpartners. \cite{batsigl, bhatta}}.   
 Both the Z-burst model and models inspired by string theory of $d>4$ use
EECR neutrinos to increase the pool of sources.

 The ``Z-burst'' models \cite{fargion} need no particle physics beyond the Standard Model
(apart from non-zero neutrino masses)
and  assume  standard hot Big Bang  cosmology.
They assert that extreme energy neutrinos (E$=M_{Z}^2/2m_{\nu}$) forming  Z bosons resonantly
with antineutrinos  of the neutrino halo  around our galaxy (or supergalaxy)
  could  produce enough protons  
well inside the GZK radius to account for the EECR. The difficulties with this model
were pointed out in ref.~\cite{sigl, domokos}. Basically, the necessary requirements on
the high energy neutrino flux and/or on the density  of halo neutrinos make the scenario
 (nearly) incompatible with other astrophysical observations.

\section{String inspired models}
  
 In all string inspired models  the neutrino nucleon cross section is above the Standard Model
value at very high energies. Internal consistency of string theories requires that strings live in a
multidimensional space (typically, d=10 for superstrings). It has been realized a few years ago
that the connection between the string scale and the  Planck scale is less rigid than hitherto
believed \cite{string}.

 In some models  with large compactified extra dimensions, (or with  a four-dimensional
brane world) together with TeV scale quantum gravity \cite{tev} EECR neutrinos
interact gravitationally with the nucleons in the atmosphere; as a matter of fact, that becomes the dominant
interaction \cite{power}. Unfortunately, the gravitational interaction does not give rise to a sufficiently
rapidly growing cross section \cite{emparan}. As a consequence, observational bounds on deep (nearly
horizontal) showers are violated and it is not possible to reach cross sections comparable to hadronic
ones around the GZK cutoff. The production of ``mini black holes'' has been also studied~\cite{emparan, feng}.
Presently, it is unclear whether the upper bounds on deep showers are compatible with these theories.

 At present, there is no internally consistent, phenomenologically viable string model known,
in which even the basic features of the dynamics -- including a mechanism of compactification --
would be satisfactorily understood. Nevertheless, various string models have so many attractive
properties that one is tempted to abstract their robust features and see whether some reasonable
conjectures can be made once     CMS energies of the colliding particles reach the string 
scale~\cite{domokos,skd}.
 For the sake of
argument, let us have a string scale of the order of 50 - 100 TeV in mind. This can be reached in
ultra high energy cosmic ray interactions: for instance, the ``gold plated'' Fly's Eye event
mentioned before has about
500~TeV in the CMS. It was shown by means of an explicit calculation~\cite{cornet,emparan}
that weakly coupled string models cannot explain the trans-GZK cosmic ray interactions. Since we
cannot calculate within the framework of a strongly coupled theory, we use features of current models,
which are likely to be present in future, phenomenologically successful theories. 
 The following basic ingredients are used:
\begin{itemize}
\item Unitarity of the $S$-matrix.
\item A  rapidly rising level density of resonances in dual models.
\item Unification of interactions at around the string scale, hereafter denoted by $M$.
\item Duality between resonances in a given channel and Regge exchanges in crossed channels.
\end{itemize}
Since  duality between resonances
and Regge {\em poles} is exact only in the tree approximation to a string amplitude,
it is unclear what the precise form of a generalization to world sheets of higher genus is:
probably, resonances of finite width are dual to Regge cuts. Thus, our formul{\ae}  are
likely to be valid to logarithmic accuracy.
Using the optical theorem,  the total cross section 
for the neutrino-parton interaction  is\footnote{All energies are assumed to be large compared to
the rest energies of the incoming particles.}:
\begin{equation}
\hat{\sigma} ( \hat{s} ) = \frac{8\pi}{\hat{s}}\sum_j^{N \left( \hat{s} \right)}
(2j + 1) \left( 1- \eta_j \cos \left( 2\delta_j \right)\right),
\label{partonsigma}
\end{equation}
where, as usual, $\eta$ and $\delta$ stand for the elasticity coefficient and phase shift
of a given partial wave, respectively. The quantity $N\left( \hat{s} \right)$ is the level of
the resonance, equal to the maximal angular momentum. 
For elastic resonances, $\eta = 1$ and $\delta \approx \pi /2$ within the width 
of the resonance. Close to the threshold of the string regime,
{\em on resonance} the total cross section is just proportional
to the number of states at a given level.  In any
realistic model the resonances have  finite widths,
thus  we  average the cross section over an energy interval
comparable to the widths of the resonances. In such an approximation,
the {\em density of states}, $d\left( \hat{s} \right)$ can be introduced
 and  $N$ is regarded as  a continuous variable,
such that $N \approx \hat{s} /M$\footnote{In the last formula, the Regge intercept
has been neglected. However, we shall see shortly that the excitations begin to
contribute significantly to the cross section for $N\geq 10$ or so; hence this approximation
is justified.}.  Using this, one gets from eq.~(\ref{partonsigma}):
\begin{equation} 
\hat{\sigma} \approx \frac{16\pi}{\hat{s}} d\left(\hat{s} \right). 
\label{lowexcit}
\end{equation}
 As inelastic channels open up, the elasticity  coefficients in eq.~(\ref{partonsigma})
become less than unity and eq.~(\ref{lowexcit}) is no longer valid. Without any detailed 
knowledge of the inelastic channels (world sheets of a higher genus in present day 
string models), we can estimate the behavior of the cross section as 
$\hat{s} \rightarrow \infty$ only. Duality tells us that the leptoquark excitations should be dual
to the exchange of the $Z$-trajectory in the $t$-channel. 
Hence, apart from logarithmic corrections,
\begin{equation}
\hat{\sigma} \sim \hat{s}^{(\alpha (0) -1)},
\label{asympt}
\end{equation}
where $\alpha (0) $ is the intercept (branch point, respectively) of the $Z$ trajectory.
Apart from corrections of the order of $(M_{Z}/M)^{2}$, one has $\alpha (0) =1 $, so that
the neutrino-parton cross section tends to a constant. (We  verify {\em a posteriori}
that $M_{Z}/M \ll 1$, so that the power corrections to the cross section are insignificant
at all energies of interest.)

The level density is a rapidly rising function of $\hat{s} $. It is known that asymptotically
it rises as $\exp (a \sqrt{\hat{s}  /M})$, with $a$ being some constant;
 see, for instance~\cite{greenetal}. However, the rise is more rapid at the beginning of the spectrum.
The first few levels of the open superstring can be well interpolated by 
the function
\begin{equation}
d(N) \propto  \exp 1.24 N, \qquad N \approx \hat{s} /M ,
\label{density_of_states}
\end{equation}
(This approximate formula was calculated from the generating function of the level density, \cite{greenetal}
eq.~(4.3.64).) General considerations on a strongly coupled string model take us this far.
To connect the low excitation regime, eq~(\ref{lowexcit}) and the 
asymptotic one, eq.~(\ref{asympt}) we chose an interpolating function guided by merely the requirement of 
simplicity. We found that 
after averaging over the parton distribution within the nucleon, the results are
insensitive to the detailed form of  the $\nu$-quark cross section. For that reason, we
chose a simple form satisfying the limits at low and high excitations:
\begin{equation}
\hat{\sigma} = \Theta \left(\hat{s}-M^{2}\right)  \frac{16\pi}{M^{2}}\frac{40 \exp 1.24 N_{0}}{1 +
 \frac{\hat{s}}{M^{2}}\exp 1.24 \left( N_{0} - \hat{s}/M^{2}\right)}
\label{interpolate}
\end{equation}

In eq.~(\ref{interpolate}), $M$ is the string scale and $N_{0}$ is a parameter measuring
the onset of the ``new physics''. In fact, one can convert that dimensionless parameter
into an energy scale. Using our previous relations, one can write $N_{0}\approx \hat{s_{0}}/M$,
or in terms of a laboratory energy of the incoming neutrino, $N_{0}\approx 2m\hat{E_{0}}/M$,
$m$ being the mass of the nucleon. In all these equations, the ``hat'' over the energies
 involved serves as a reminder that the quantities have to be integrated  over the parton
distribution. As usual, the conversion is carried out by means of  substitutions such as
$\hat{s}= x s$, $x$ being the momentum fraction of a parton within the nucleon. The step
function is inserted because the cross section of the ``new physics'' vanishes at CM
energies below the mass of the first resonance. The parton distributions have been taken from
CTEQ6 \cite{cteq6}. The dominant contribution comes from valence quarks; gluons
do not contribute, since no presently known unification scheme contains ``leptogluons''.
Finally, the contribution of the sea is negligibly small, since the latter is concentrated
around $x=0$.

 It is impossible to precisely determine the two parameters, $M$ and $N_0$ entering
eq.~(\ref{interpolate}).  Nevertheless, the parameters can be bounded by the trans-GZK data.
As it was mentioned before
\begin{itemize}
\item no deep showers have been observed by AGASA and Fly's Eye,
\item the trans-GZK showers appear to be
``hadron-like'', {\em i.e.} they originate high in the atmosphere and appear
to exhibit a development resembling proton induced showers.
\end{itemize}
These constraints were analysed  \cite{gsigl}.  The absence of deep showers excludes a region of the 
neutrino cross section, approximately, 0.02~mb $\leq \sigma \leq $ 1~mb. The cross section
has to grow fast to roughly hadronic size around the ``ankle'' in the cosmic ray spectrum,
 approximately at $5\times 10^{19}$~eV
and stay of this size or grow slightly. Unless these conditions are satisfied, the neutrino
model of trans-GZK cosmic rays fails. 
A search of the parameter space yields
 $E_{0}\approx 5\times 10^{10}$~GeV and $M\approx 80$~TeV. With these values of $M$ and $N_0$
the  cross section  is rising sufficiently rapidly to satisfy the deep shower bound
and at the same time, it is  sufficiently large in the trans-GZK energy
region. These values of $E_{0}$ and $M$ give $N_{0}\approx 15.6$ confirming the intuitive
expectation. The neutrino-nucleon cross section  is shown
in Fig.~(\ref{crossection}).
\EPSFIGURE{
 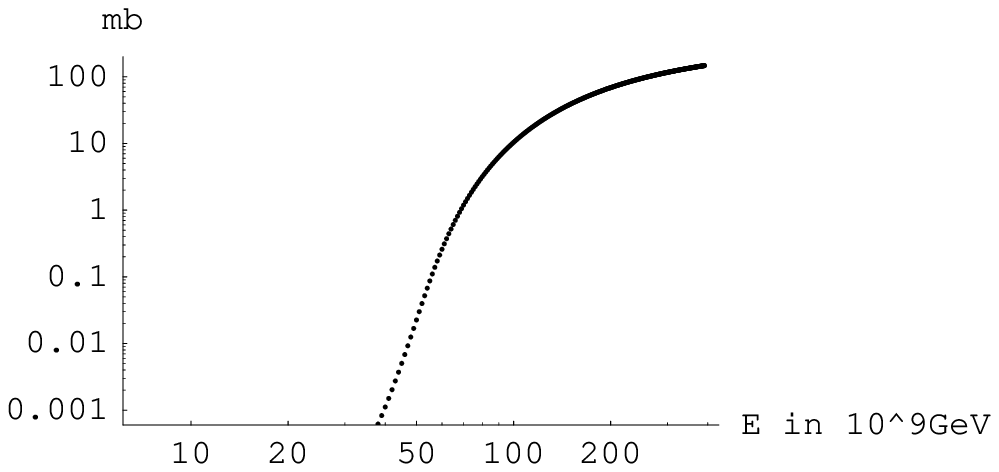}{The $\nu$-nucleon cross section calculated from eq.~(\ref{interpolate}) and
the CTEQ6 parton distribution. $E_{0} = 5\times 10^{19}$eV, $M= 80$TeV.\label{crossection}}

Due to the exponential dependence of eq.~(\ref{interpolate})
on the parameters, one cannot vary their values over a broad range without getting a
contradiction either with the bound on deep showers and/or with the required value
of the cross section for trans-GZK showers.
Neutrino induced showers were simulated using the ALPS (Adaptive Longitudinal Profile
Simulation) Monte Carlo package authored by Paul~T. Mikulski\cite{pault}. Similarly to
earlier studies, see, {\em e.g.}~\cite{jhep} it was assumed that quarks and leptons are 
created in comparable numbers in an interaction as long as the CM energy of an interaction
 remains above 
$M$. Once the energy drops below $M$, the usual Standard Model cross sections
govern the further development of the shower. A qualitative consequence of this feature
is that, statistically, neutrino induced showers exhibit larger fluctuations than
proton induced ones, see~\cite{jhep}\footnote{This is a consequence of the fact that,
once in the standard model regime, leptonic interactions have a lower average multiplicity
than hadronic ones.}.  In addition, once the cross section becomes larger than about 15-20 mb,
the shower starts high in the atmosphere. Hence, on an event by event basis, 
such showers are virtually indistinguishable from hadron induced showers

\section{Conclusions}
 
The trans-GZK energy region presents a unique opportunity to get hints on the qualitative
features of  theories going beyond the Standard Model. In astrophysics, these extensive air showers may
bring information either about cosmologically early times or about the most violent regions in 
the Universe. To be  able to distinguish among these numerous imaginative scenarios, the collection 
of high statistics data is necessary. In the near future this will become possible with 
the continued operation and improvements of AGASA and HiRes and with the start of the hybrid (grand
array and air fluorescence) megadetactor, the Pierre Auger Observatory.

\section{Acknowledgement} We thank Professors K.
 Meier and O. Nachtmann for the organization and
running of a very stimulating and inspiring Workshop.

\end{document}